# Polarisation measurements with a CdTe pixel array detector for Laue hard X-ray focusing telescopes

Ezio Caroli[1,*], Rui M. Curado da Silva[2], Alessandro Pisa[3], John B. Stephen[1], Filippo Frontera[3,1], Matilde T. D. Castanheira[2], Stefano del Sordo[4]

[1]*INAF/IASF-Bologna, Via Gobetti 101, 40129 Bologna, Italy*
[2]*Departemento de Fisica da Universidade de Coimbra, 3000 Coimbra, Portugal*
[3]*Dip. di Fisica, Università di Ferrara, Via Saragat 1, 44100, Ferrara, Italy*
[4]*INAF/IASF-Palermo, Via Ugo La Malfa 153, 90146 Palermo, Italy*
(* author for correspondence , e-mail: caroli@iasfbo.inaf.it)

**Abstract.** Polarimetry is an area of high energy astrophysics which is still relatively unexplored, even though it is recognized that this type of measurement could drastically increase our knowledge of the physics and geometry of high energy sources. For this reason, in the context of the design of a Gamma-Ray Imager based on new hard-X and soft gamma ray focusing optics for the next ESA Cosmic Vision call for proposals (Cosmic Vision 2015-2025), it is important that this capability should be implemented in the principal on-board instrumentation. For the particular case of wide band-pass Laue optics we propose a focal plane based on a thick pixelated CdTe detector operating with high efficiency between 60-600 keV. The high segmentation of this type of detector (1-2 mm pixel size) and the good energy resolution (a few keV FWHM at 500 keV) will allow high sensitivity polarisation measurements (a few % for a 10 mCrab source in $10^6$s) to be performed. We have evaluated the modulation Q factors and minimum detectable polarisation through the use of Monte Carlo simulations (based on the GEANT 4 toolkit) for on and off-axis sources with power law emission spectra using the point spread function of a Laue lens in a feasible configuration.

## 1. Introduction

Polarimetry in astrophysics is still a quite unexplored domain in the X- and γ-ray energy band. By measuring the polarization degree and orientation of the radiation emitted by an astronomical source, a new, important observational parameter is obtained in addition to spectral and time variability information. Polarimetric observations offer novel information about constitution, physical processes and emissions of various sources, e.g. Pulsars, Solar Flares, Active Galactic Nuclei, Galactic Black Holes or Gamma-Ray Bursts [Lei et al., 1997]. However, to date in the energy domain from hard X-rays up to soft γ-rays no dedicated polarimeters have ever been launched into space or on a balloon-borne experiment. Indeed for those few instruments which are capable of performing this type of measurement, polarimetry itself plays a secondary role in the mission objectives, the main purpose of these instruments being to perform spectral imaging and timing analysis of celestial sources. This is the case for COMPTEL (COMPton TELescope) [Schönfelder et al., 1993] instrument on board CGRO (Compton Gamma-ray Observatory), RHESSI (Reuven Ramaty High Energy Solar Spectroscopic Imager) [McConnell et al., 2002] and IBIS (Imager on Board the INTEGRAL Satellite), that is part of the INTEGRAL (INTErnational Gamma-Ray Astrophysics Laboratory) mission [Ubertini et al., 2003].

Several years ago, a collaboration between INAF/IASF – Bologna (Istituto di Astrofisica Spaziale e Fisica Cosmica), Italy and the Departamento de Física da Universidade de Coimbra, Portugal started with the aim of studying the performance of solid state (in particular CdTe/CZT) detector arrays as high-energy (above 100 keV) Compton polarimeters. The main advantage of this kind of detector is that high-sensitivity polarimetric measurements can be performed simultaneously with spectral imaging and time variability observations. The collaboration has already proposed an instrument concept to measure the Crab pulsar polarization level in the range between 50-1000 keV: the CIPHER (Coded Imager and Polarimeter for High Energy Radiation) telescope [Curado et al., 2003]: a Minimum Detectable Polarisation (MDP at 3σ) level below 10% is expected for a typical balloon observation time ($10^4$ s).

Because of the scientific importance of polarimetry measurements it is the right strategy to implement this capability in the principal on-board instrumentation in the design of a Gamma-Ray Imager based on new hard-X and soft gamma ray focusing optics for the next ESA Cosmic Vision call for proposals (Cosmic Vision 2015-2025), it is strategic to implement this capability in the principal on-board instrumentation. Herein we present a preliminary study of the achievable performance in high energy polarimetry for a Hard X-ray focusing Telescope based on wide band-pass Laue optics [Pisa et al., 2005] with a focal plane based on a thick pixelated CdTe/CZT detector operating with high efficiency between 60-600 keV.

## 2. Scattering polarimetry: numerical model and experimental confirmation

The GEANT4 package is a very suitable and efficient tool with which to implement a Monte Carlo code that simulates the response of a CdTe detection plane designed for polarimetric measurements. The code is logically split into two main tools: (a) the modules implementing the physics of the electromagnetic interactions of polarized photons in particular for the Compton scattering; (b) the detection system with the definition of the beam characteristics, the detection plane design (geometry and material) and the read-out logic.

### 2.1. Compton scattering polarimetry model

Polarimetric measurements are based on the fundamental concepts associated with polarised Compton interactions. In the X- and γ-ray energy range, this electromagnetic interaction can be the cause of a non-uniformity in the spatial distribution of the detected photons. After undergoing Compton scattering, the polarised photons' new direction depends on the orientation of its polarisation vector before the interaction. If photons interact at least twice inside the detector, the position of the first and the second interaction (double event) allow us to infer the polarisation degree and direction of the radiation beam. The Klein-Nishina cross-section for linearly polarised photons gives an azimuthal dependency for the scattered photons:

$$\frac{d\sigma}{d\Omega} = \frac{r_0^2}{2}\left(\frac{E'}{E}\right)^2\left[\frac{E'}{E}+\frac{E}{E'}-2\sin^2\theta\cos^2\varphi\right], \quad (1)$$

where $r_0$ is the classical electron radius, $E$ and $E'$ are the energies of the incoming and outgoing photons respectively, $\theta$ is the angle of the scattered photon and $\varphi$ is the angle between the scattering plane (defined by the incoming and outgoing photon directions) and incident polarisation plane (defined by the polarisation vector and the direction of the incoming photon). As can be seen from (1), fixing all other parameters the scattering probability varies with the azimuthal angle $\varphi$. The maximum relative difference between the cross-section values arises for $\varphi = 0°$ – where the cross-section reaches a minimum – and $\varphi = 90°$ – where it reaches a maximum. This relative difference is maximised for a scatter angle $\theta_M$, which depends on the incident photon energy. For soft gamma and hard X-rays the $\theta_M$ value is always around 90° [Lei et al., 1997]. A planar detection plane configuration is very suitable to polarimetric measurements since the planar configuration fits an essential condition: after a Compton interaction, outcoming photons with low and high $\theta$ values – when polarimetric asymmetries are less evident – take a direction away from the detection plane, while those emerging from interactions with θ close to 90° have a stronger probability of going through a second interaction inside the detector. Thus with a rather simple design, such as planar detector, it is possible to obtain a double events distribution that allow us to perform polarimetric studies of celestial bodies or phenomena.

## 2.2. Experimental verification of the MC model

The collaboration between the two research groups has already given rise to both extensive Monte Carlo (MC) studies [Curado et al., 2003] and experimental tests of pixel CdTe matrices used as polarimeters that have confirmed the reliability of our MC codes. These Monte Carlo simulations allowed both the study of CdTe polarimeters (Q factor and detection efficiency) and the study of several problems and effects associated with the interaction of polarized radiation in planar pixel CdTe detectors, such as the influence on polarimetric performance of the X- and γ-ray incidence inclination with respect to the optical axis, full irradiated detector surface analysis techniques and the variation of modulation factor as a function of the angle between the polarisation direction and the detection plane axis. Experimental polarimetric studies were performed at the European Synchrotron Radiation Facility in Grenoble, France, where some 4×4 CZT pixelized matrices (with various thicknesses) were tested with a ~100 % polarized photon beam [Caroli et al., 2003] at different energies between 100 and 500 keV. These tests have shown a good agreement between the results obtained from our MC simulation model with the geometry adapted to the experimental set-up and the beam measurements. The differences between the values of the $Q$ factor obtained by MC simulation and by beam tests are small (5-10%) and within the statistical errors [Curado et al., 2004]. These results have confirmed the correctness of the numerical Compton scattering model that was developed by some of the authors and that was included in the GEANT4 code since the version 4.3. The conclusions obtained from these studies are very important in order to develop and to correct Laue lens CdTe/CZT focal plane polarimetric studies.

# 3. Laue lens telescope model geometry

The development of focusing optics is necessary in order to increase the statistical quality of the X- and gamma-ray spectra of the celestial sources and to improve the study of the emission mechanisms of the celestial sources, their nature, their evolution with time and their contribution to the Cosmic X-ray Background. Recently the development of high energy Laue lenses with broad energy band-passes from 60 to 600 keV has been proposed for new high sensitivity hard X

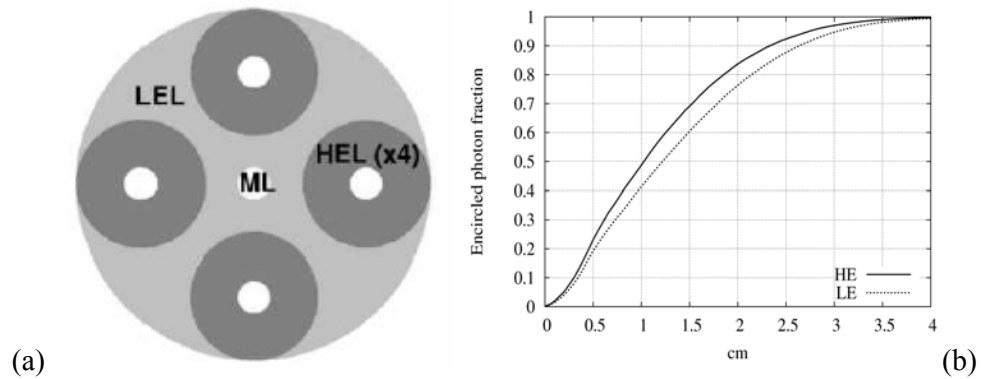

(a) (b)

Figure 1. (a) a possible multi lens arrangement, where the LEL energy band-pass extends from 60 keV up to 200 keV and the HEL band-pass ranges from 150 keV up to 600 keV; (b) the focused photon percentage profile integrated over the expected polychromatic PSF for an on-axis source in HE and LE lens with a focal length of 50 m.

ray telescopes [Frontera et al., 2006] in multi–lens configurations.

The configuration of the multi-lens telescope (Fig.1a) includes a Low Energy Lens (LEL, light grey circle) with a nominal passband from 60 to 200 keV, 4 High Energy Lenses (HEL, dark grey circles) with a nominal pass-band from 150 to 600 keV and in the cavity left free from the LEL, a Wolter I Multi-Layer (ML) confocal optics covering the 10-80 keV energy band, that will not be considered in our studies. The focal length baseline is about 50 m. The focal plane configuration proposed is based on a thick CdTe pixel detector in order to optimise the detection efficiency at high-energy (from 60 to 600 keV). A CdTe thickness of at least 3 mm (better 10 mm) is required, with pixel scales of the order of a few millimetres since the expected point spread function (PSF), for a 50 m focal length and for an on-axis source, has the integrated profile of Fig. 1b where 95% of the focussed photons lay inside a circular surface of 30 mm in diameter [Pisa, et al., 2006].

To be compatible with this point spread function, a matrix composed of 32×32 CdTe pixels was simulated, where each pixel has a 2×2 mm$^2$ surface area. Each pixel of the focal plane detector is labelled by an identification number (*x pixel ID*, *y pixel ID*) along the *x*- and *y*-axis parallel to the detector surface. The photon irradiation geometry implemented in the code allows the simulation of the near-uniform distribution of Laue focussed photons over an area as large as the expected PSF. Then a read out geometry identifies the pixels where double and higher order events occur, counting the number of these events in each pixel pair, as well as the respective energy deposition.

Since X-ray focusing with Laue lenses is based on low incidence angle diffraction, it has a negligible effect on the initial photons' polarization state for high energy photons (at least above 10-20 keV). Therefore, throughout this study it was considered that Laue lenses had no affect on the polarization state of X- and gamma-ray celestial emissions [Frontera et al., 1994].

## 4. Focal plane Polarimetric performance

Both the modulation *Q* factor and the minimum detectable polarisation of the CdTe focal plane have been evaluated by GEANT4 based Monte Carlo simulations for a Crab pulsar type spectrum and using as expected background a value derived by a geometrical scaling (according to detector volume and FOV) from that given in [Curado da Silva, et al., 2006] for the LOBSTER CZT Gamma Ray Burst monitor and considering only the analytical response (PSF) of the Laue lens [Pisa et al., 2005].

### 4.1. Polarimetric Q factor and detection efficiencies

The polarimetric performance of an instrument can be evaluated by analysing the distribution of double events through the modulation factor, *Q*. This is obtained by integrating the Compton differential cross section formula over the solid angles defined by the physical geometry of the detection. For a pixel detector *Q* can be expressed as [Lei et al., 1997]:

$$Q = \frac{N_{//} - N_{\perp}}{N_{//} + N_{\perp}}, \qquad (2)$$

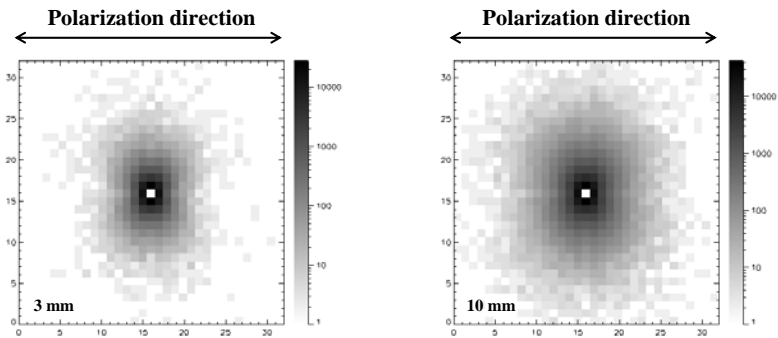

Figure 2 – Double event distribution maps (only second interactions) obtained by Monte Carlo simulations for a 3 mm and a 10 mm thick focal plane, when irradiating a detector central pixel by a Crab-type 100 % polarized photon beam focused by the LEL in the energy band 120-200 keV.

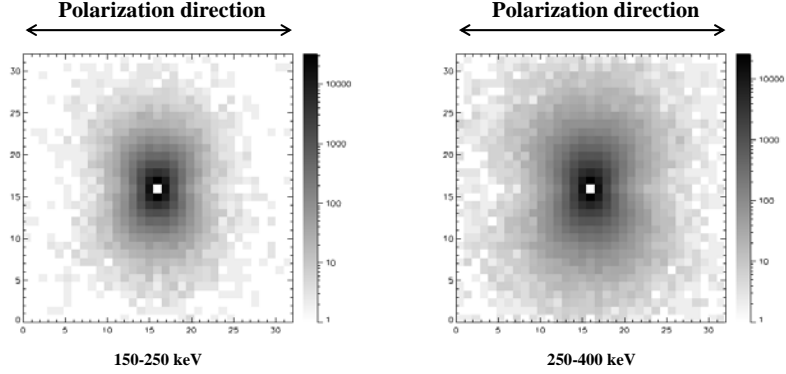

Figure 3 – Double event distribution maps obtained by Monte Carlo simulations in the 120-250 keV and 250-400 keV energy bands, when irradiating a 3 mm thick focal plane by a Crab-type 100 % polarized photon beam focused by the HEL

where $N_{//}$ and $N_{\perp}$ are the number of counts integrated in two orthogonal directions over the detection plane.

In the simulations, a surface equivalent to the PSF area containing 95% of focused photons, was irradiated uniformly by a 100 % linearly polarised beam of $10^7$ photons, with a Crab-like power law energy spectrum ($\propto E^{-2}$ photons/cm$^2 \cdot$s$^{-1} \cdot$keV$^{-1}$). The beams' energy was divided into intervals corresponding to energy bands where the lenses have a nearly uniform response: 60-120 keV and 120-200 keV for LEL; 150-250 keV, 250-400 keV and 400-600 keV for HEL.

Using the position of the two hits of each double (scattered) event, distribution maps (like those in Fig. 2 and 3) are built where the pixel of one of the two hits is normalized to the centre and the other is added in the relative position in the map. At the end, each element of the double events distribution maps contains the number of second interactions from each scattering pixel of the detection plane and gives information on the angular distribution of Compton scattered events. As expected, an asymmetry arises in the double event distribution maps due to the polarization of the incoming beam. As can be seen, this asymmetry is perpendicular to the direction of the vector electric field of the beam, as was also expected. As shown in Fig. 2, the higher detection efficiency of the 10 mm thick detection plane results in a higher detection rate of Compton double events. The double events distribution maps represented in Fig. 3 were obtained by photons focussed by one of the High Energy Lenses in two different energy bands: 150 keV to 250 keV and 250 keV to 400 keV. In the 250 to 400 keV energy range there is a higher probability that a Compton photon escapes the first interaction pixel and produces a second interaction in a neighbor pixel, which explains the higher number of double events detected compared with 150 to 250 keV energy range.

The double event distributions obtained for a combination of Laue lens type energy band-pass and the focal plane thicknesses of 3 mm (inferior limit) and 10 mm (superior limit) allowed us to calculate the corresponding $Q$ factors from (2) and the respective double event absolute efficiencies (double events detected/total number of photons focussed by the lenses) integrated over each lens energy band-pass.

Table I - $Q$ factors and double events absolute efficiencies obtained for HE and LE Laue lenses energy bands, considering a 3 mm (a) and a 10 mm (b) thicknesses focal plane.

| | Low Energy Lens | | | | High Energy Lenses | | | | | |
|---|---|---|---|---|---|---|---|---|---|---|
| *ΔE (keV)* | 60-120 | | 120-200 | | 150-250 | | 250-400 | | 400-600 | |
| *Thickness (mm)* | 3 | 10 | 3 | 10 | 3 | 10 | 3 | 10 | 3 | 10 |
| *Q factor* | 0.30 | 0.29 | 0.35 | 0.32 | 0.36 | 0.33 | 0.37 | 0.33 | 0.36 | 0.31 |
| *Absolute Efficiency (%)* | 0.11 | 0.13 | 1.43 | 2.86 | 2.35 | 5.63 | 3.57 | 12.3 | 2.82 | 13.0 |

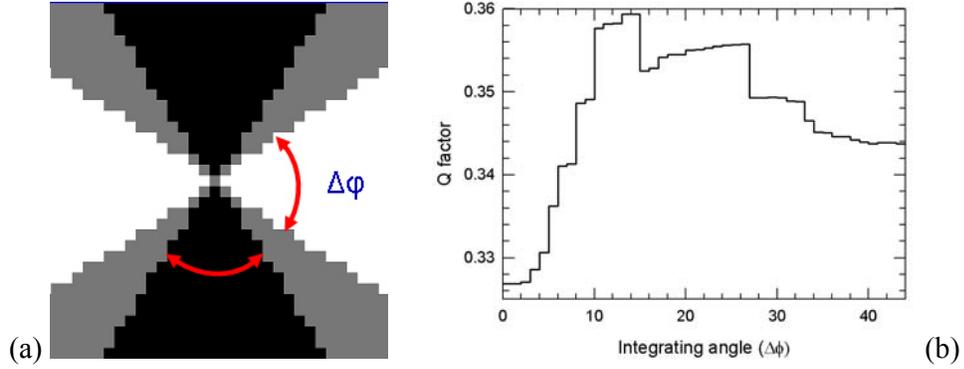

Figure 4. The optimization of the Q factor: (a) a graphical representation of a method to improve the Q factor integrating over two angular sector with varying aperture (Δφ) around the two orthogonal direction corresponding to the detector sides; (b) the modulation Q factor as a function of Δφ, showing that the best value is achievable considering 10-15 degree wide detector sector.

Table I shows that for each Laue Lens type the Q factor is always higher for a 3 mm than a 10 mm thick focal plane, when the same energy range is considered. This is because for 3 mm thickness the fraction of double events detected that undergo Compton interactions whose $\theta$ is very close to 90º, is higher than the same fraction of double events detected for a 10 mm pixel thickness focal plane. For a larger pixel thickness the scattering acceptance angle is wider and therefore will decrease on average per-photon information about the polarization status of the scattered photons, which is optimal when this angle is close 90°. The decrease of the $Q$ factor caused by Compton interactions whose $\theta$ diverge from 90° also occurs for photon energies higher than 400 keV, however in this case it is due to the Compton scattering physics. The mean Compton scattering angle decreases as the photon energy increases and therefore the number of double events occurring at polar angles around 90º is lower. As shown in previous Monte Carlo simulation studies performed over a 100 keV to 1 MeV energy range this effect becomes important for photon energies higher then 400 keV [Curado et al., 2003].

From Table I it is evident that the double event absolute efficiency obtained is higher for thicker focal planes under the same simulation conditions and it increases with the photons' energy, since the probability that the second interaction occurs outside the pixel where the first interaction took place increases as well.

Different methods can be applied to the original double event distribution to optimize the evaluation of the Q factor. A simple procedure is based on the fact that there is no *a-priori* restriction on the linear distance between the point of interaction inside the detector plane for valid events, thus allowing optimisation of Q by defining the acceptance areas of events to be included in the sums of $N_\perp$ and $N_\parallel$ depending as shown in Fig. 4a. The Q evaluated with this event selection as a function of the acceptance angle is given in Fig. 4b. A further improvement in the Q evaluation can be obtained by combining the event acceptance angular opening with different scattering distance intervals (Stephen et al., 2001). A proper evaluation (*effective Q*) should include the weighing of the calculated Q with its statistical significance that depend on the fraction of double events (N(Δφ)/$N_{tot}$) considered in the selection:

$$Q_{effective} = Q(\triangle\varphi) \cdot \sqrt{\frac{N(\triangle\varphi)}{N_{tot}}}, \quad (3)$$

In fact the Q increases with linear separation between the two hits of a scattered event, while the efficiency decreases, leading to a maximum of the *effective Q* at some distance from the original interaction point, depending on the energy and on the geometrical characteristics of the detection plane (e.g. size of the pixel, thickness, etc.). In the case illustrated in Fig. 4 the $Q_{effective}$ reach the maximum value of 0.44 for an integration angle of ~30°. Greater integration angles do not further improve the polarimetric performance of the detector. The a priori knowledge of this limit could allow a reduction of background contributions (e.g. due to random coincidences).

## 4.2. Minimum Detectable Polarization

In order to get a closer approximation to the in-flight measurements, background noise was included in our model. The background noise level caused both by cosmic radiation and from secondary photons generated by particles interacting with the elements that compose the telescope can be critical to polarimetric measurements.

The minimum detectable polarisation (MDP) of a polarimeter in the presence of background noise can be expressed as [Lei et al., 1997]:

$$MDP(100\%) = \frac{n_\sigma}{A \cdot \varepsilon \cdot S_F \cdot Q_{100}} \sqrt{\frac{A \cdot \varepsilon \cdot S_F + B}{T}}, \qquad (4)$$

where $n_\sigma$ is the significance, $Q_{100}$ the modulation factor for a 100 % polarized source, $\varepsilon$ the double event detection efficiency, $A$ the polarimeter detection area in cm$^2$, $S_f$ the source flux (photons·s$^{-1}$·cm$^{-2}$), $B$ is the background flux (counts/s) and $T$ the observation time in seconds. For a $3\sigma$ detection significance we calculated the MDP expected both for LEL and HEL telescopes when on the focal plane detector is focussed a source flux with a Crab-like spectrum and in the presence of a background level. The background used is derived by geometrical scaling from that expected for the LOBSTER GRBM, based on CZT array modules, planned onboard the ISS (low earth orbit at 340 km with ~50° inclination), as described in [Amati et al., 2006]. Due to the orbital characteristics of the ISS the background has the following average spectrum: $1.5 \cdot E^{-1.4}$ photons/cm$^2$.s$^{-1}$.keV$^{-1}$. For a HAXTEL-like mission, which will operate in a very eccentric or L2 orbit, we have assumed a background normalisation of 0.3 (1/5 of the LOBSTER case) taking into account, the more efficient shielding possible for a smaller focal plane detector, and the possibility to further reduce the effective background for polarimetry using Compton kinematics. The polarimeter detection area (A) in equation (4) is determined from the Laue lens effective area, while the background (B) is integrated over an area equivalent to the PSF surface (~7 cm$^2$). Therefore for each lens energy band-pass, the MDP was calculated using the respective average lens effective area for the considered Cu Laue crystal lens arrangement and the background count rates, reported in Table II. The $Q_{100}$ factor and the absolute double event efficiency included in the MDP calculations for each of the considered energy range are those presented in Table I.

Table II. The average effective area of the LEL and HEL instruments in the used energy band, with the background count rate in the same energy bands

|  | Low Energy lens | | High Energy Lens | | |
|---|---|---|---|---|---|
| *ΔE (keV)* | 60-120 | 120-200 | 150-250 | 250-400 | 400-600 |
| *Average Effective Area (cm$^2$)* | 75 | 250 | 840 | 620 | 240 |
| *Background (counts/s)* | 0.241 | 0.139 | 0.127 | 0.096 | 0.07 |

Fig. 5 shows the calculated MDP, for a $3\sigma$ detection significance, as a function of the observation time for the high and low energy Laue lens instruments respectively, for a source of 10 mCrab with a Crab-like energy spectrum. For the High Energy Laue Lens the MDP is shown for three energy ranges and considering a 10 mm focal plane, while for the Low energy Laue lens we report the MDP in the 120-200 keV band for two focal plane thicknesses (3 and 10 mm). Fig. 4 (left) shows that it is potentially possible to measure degrees of polarization lower than 20 % for 10$^4$ s observation times and lower than 2% for 10$^6$ s, in the energy range between 150 keV and 400 keV, and polarization degrees lower than 40% for 10$^4$ s observation times and lower than 4% for 10$^6$ s, in the energy range between 400 keV and 600 keV. In the higher energy band the $Q$ factor is lower, which explains the higher MDP values. In the case of a 3 mm focal plane (not shown for simplicity), the MDP calculated was slightly higher than the 10 mm focal plane results due to the lower double event detection efficiency of the 3 mm pixels. The MDP obtained for 10$^6$ s observation times for a 3 mm CdTe focal plane was: ~3% (150-400 keV) and 7.5 % (400-600 keV). Fig. 4 shows that also the Low Energy Laue Lens instrument can achieve an MDP inferior than 40 % for 10$^4$ s observation times and inferior than 3 % for 10$^6$ s in the 120 to 200 keV band-pass. The MDP calculations for the LEL have also indicated that in the 60-120 keV band-pass polarization measurements are effective only for observation timescales above $10^5$ s: e.g. for $10^6$ s observations MDP at the level of ~27 % and ~23 % (for 3 mm and 10 mm thickness focal plane respectively) are achievable. This result is easily explained by the low double event efficiency (Table I) in the lower energy band of the LEL instrument.

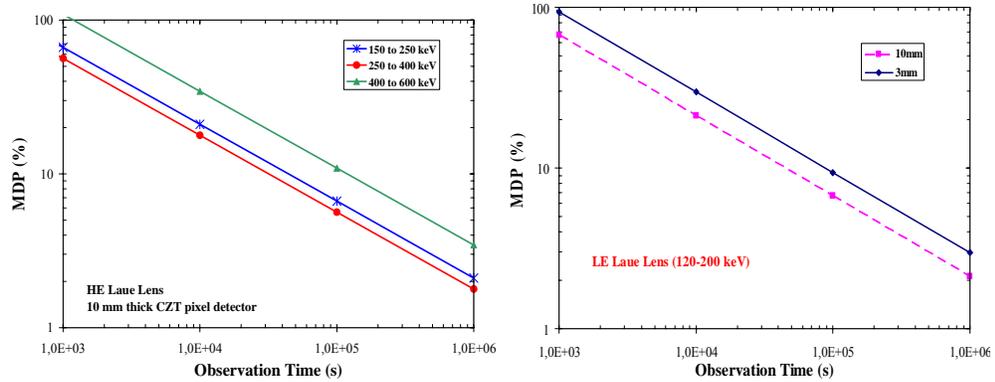

Figure 5 - (left plot) MDP in different energy bands for a 10 mm thick CdTe/CZT focal plane for the HE Laue lens telescope. (right plot) MDP in the 120-200 keV energy band for a 3 mm and a 10 mm CdTe/CZT focal plane for the LE Laue lens instrument. Both MDP were obtained for a $3\sigma$ detection significance and considering a 10 mCrab emitting source.

In order to study the optimal pixel size of the Laue lens telescope focal plane, the MDP was calculated as a function of pixel lateral size dimensions under the same irradiation conditions as explained before. Since the expected point spread function is of about 30 mm, a pixel scale of a few mm (1-3) would be enough have a good sampling from the imaging and source detection point of view. A smaller pixel scale would allow a better sensitivity to the polarized emission, but it means an increase in the focal plane complexity (a large number of channels require more complex electronics and more resources). Therefore we limited our study to pixel lateral dimensions between 0.5 mm and 2 mm. Fig. 6 shows the $Q$ factor and the MDP (for $10^6$ s observation time) obtained for the Low Energy Laue Lens in the 120 keV to 200 keV band-pass combined with a 10 mm thickness CdTe focal plane. Since it is constantly simulated a 32×32 CdTe matrix, for lateral pixel sizes smaller than 1.0 mm its volume is smaller and therefore a fraction of Compton photons escape from the detection plane before having a second interaction, which explains that $Q$ decreases for lateral sizes smaller than 1.0 mm. From 1.0 mm on, this effect becomes residual and smaller lateral dimensions result in higher $Q$ factors due to a higher rate of second interactions occurring inside pixels further from the central pixel, which contributes to an improvement in the double event distribution resolution. However, the net gain observed in the MDP for pixel dimensions lower than 1 mm does not compensate for the technical difficulties associated with its production. Therefore, focal plane pixel lateral dimensions of about 1 mm up to 2 mm provide a good trade off between focal plane complexity and polarimetric performance. Furthermore, we point out that the improvement in the MDP achievable with the smaller pixel scale might be obtained with less expensive background reduction techniques such as optimizing the shielding and/or applying event selection procedures based on Compton kinematics.

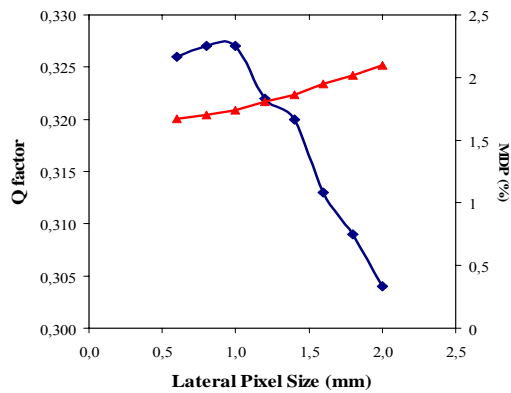

Figure 6 – The $Q$ factor (diamonds) and the MDP for a $10^6$ s observation time (triangles) in function of pixel lateral size obtained for the LE Laue Lens and a 10 mm thickness CdTe focal plane in the 120 keV to 200 keV energy band.

### 4.3. Sensitivity to polarized sources

Using the polarimetric performance obtained for the multi Laue lens configuration that we have considered we can also make a first estimate of the sensitivity to the polarization level of a source. This figure will allow to know which is the minimum detectable flux for a source with a given linear polarization level. The following equation represents, as a first approximation, the polarization sensitivity [Lei, e al., 1997]:

$$F_{pol\%} = \frac{F_{min}}{Q_{pol\%}} \ ph/(s \cdot cm^2 \cdot keV), \qquad (5)$$

in which $Q_{pol\%}$ is the Q factor for a source with a given polarized level of polarization in %. A rough estimate of this value can be obtained simply by multiplying the $Q_{100}$ for the polarized fraction of the source (*f*). In this case we have used an average $Q_{100}$=0.35 in the 60-600 keV band. For the 3σ sensitivity ($F_{min}$) to non-polarized flux we have adopted the value of $3\times10^{-8}$ photons/keV/cm$^2$/s (60-600 keV energy band, $10^6$ second observation time) as given in [Frontera et al., 2006], obtained using the HAXTEL effective area, a background level of $5\times10^{-5}$ counts/s/keV/cm$^2$ (i.e. the average of the BeppoSAX/PDS instrument between in the 60-300 keV energy range), and a unitary detection efficiency.

Table II. Evaluation of the sensitivity to the polarization level of a source

| Source Polarisation Level | 3σ Sensitivity for a polarized source [Ph/cm2/s/keV] For T=$10^6$ s and ΔE = 60-600 keV | |
|---|---|---|
| 50% (*f=0.5*) | $1.7\times10^{-7}$ | (~0.6 mCrab) |
| 10% (*f=0.1*) | $8.5\times10^{-7}$ | (~3 mCrab) |
| 2% (*f=0.05*) | $4.3\times10^{-6}$ | (~15 mCrab) |

## 5. Conclusion

The MC polarimetric results obtained for a wide band Laue lens instrument show that a pixellised CdTe focal plane is potentially well suited to the task of performing very sensitive polarimetric measurements. For typical satellite observation times (up to $10^6$s), polarization down to a level as low as 2% can be measured for X- and gamma-ray emissions from a 10 mCrab source. The Focal plane pixel dimensions study showed that a lateral pixel dimension of 1-2 milllimetres provides optimal polarimetric performance and that a thicker focal plane (10 mm) is a better choice.

*Acknowledgment*

This work was carried out in cooperation between the Departamento de Física, Universidade de Coimbra, Portugal (Unit 217/94) and the IASF – Sezione di Bologna (*Istituto di Astrofisica Spaziale e Fisica Cosmica*), CNR, Italy and was supported in part by FEDER through project POCTI/FP/FNU/50228/2003 of *Fundação para a Ciência e a Tecnologia*, Portugal. The work of R.M. Curado da Silva was supported by the Fundação para a Ciência e Tecnologia, Portugal, through the research grant SFRH/BPD/11670/2002.